\documentstyle[12pt,a4,epsfig]{article}

\begin{document}
\newcommand {\epem} {e$^+$e$^-$}
\newcommand {\nthreejet} { {\mathrm{N}}_{\mathrm{3-jet}} }
\newcommand {\nzerog} { {\mathrm{N}}^0_{\mathrm{g}} }
\newcommand {\cff} {\mathrm{C}_{\mathrm F}}
\newcommand {\caa} {\mathrm{C}_{\mathrm A}}
\newcommand {\cacf} {\mathrm{C}_{\mathrm A}/\mathrm{C}_{\mathrm F}}
\newcommand {\nee}  { {\mathrm{N_{e^+e^-}}}  }
\newcommand {\twoqscale} { 2E^{*} }
\newcommand {\qscale} { E^{*} }
\newcommand {\gscale} { p_{\perp} }
\newcommand {\qq} {$q\overline{q}$}
\newcommand {\qqg} {$q\overline{q}g$}
\newcommand {\ecm} {$E_{\,\mathrm{c.m.}}$}
\newcommand {\lmsbar} {\Lambda_{\overline{\mathrm{MS}}}}
\newcommand {\rch} {r_{ch.}}

\newcommand {\pp} {p$\overline{\mathrm{p}}$}
\newcommand {\durham} {$k_\perp$}
\newcommand {\pperp} {$p_\perp$}
\newcommand {\ntotal} {$n_{\mathrm{total}}$}
\newcommand {\nch} {$n_{\mathrm{ch.}}$}
\newcommand {\mnch} {$\langle n_{\mathrm{ch.}}\rangle$}
\newcommand {\mngl} {$\langle n \rangle_{\mathrm{gluon}}$}
\newcommand {\mnqu} {$\langle n \rangle_{\mathrm{quark}}$}
\newcommand {\mnratio} {$\langle n \rangle_{\mathrm{gluon}}/
                        \langle n \rangle_{\mathrm{quark}}$}
\newcommand {\lms} {$\Lambda_{\overline{\mathrm{MS}}}$}
\newcommand {\nparton} {n_{\,\mathrm{parton}}}
\newcommand {\nhadron} {n_{\,\mathrm{ch.}}}
\newcommand {\gincl} {g_{\,\mathrm{incl.}}}
\newcommand {\ggincl} {(gg)_{\,\mathrm{incl.}}}
\newcommand {\thmax} {$\theta_{\,\mathrm{max.}}$}
\newcommand {\ejet} {E_{\,\mathrm{jet}}}
\newcommand {\mngincl} {\langle n_{\mathrm{ch.}} 
                         \rangle_{g_{\,\mathrm{incl.}}}}
\newcommand {\megincl} {$\langle E \rangle_{g_{\,\mathrm{incl.}}}$}
\newcommand {\egincl} {$E_{g_{\,\mathrm{incl.}}}$}
\newcommand {\mnchgl} {\langle n_{\,\mathrm{ch.}} \rangle_{\mathrm{gluon}}}
\newcommand {\mnchqu} {\langle n_{\,\mathrm{ch.}} \rangle_{\mathrm{quark}}}
\begin{titlepage}
\bigskip
\begin{center}{\large  UNIVERSITY OF CALIFORNIA, RIVERSIDE
}\end{center}\bigskip
\begin{flushright}
  UCRHEP-E265 \\
  November 9, 1999
\end{flushright}
\bigskip\bigskip\medskip
\begin{center}{\LARGE\bf 
Determination of the QCD color factor ratio
{\boldmath{$\cacf$}} from the scale dependence of multiplicity
in three jet events
}\end{center}
\begin{center}{\large J. William Gary \\[4mm]
{\it Department of Physics, 
University of California, \\[1mm]
Riverside, CA 92521 USA }
}\end{center}
\begin{center}{\Large\bf  Abstract}\end{center}

\noindent
I examine the determination of the QCD color factor ratio 
{$\cacf$} from the scale evolution
of particle multiplicity in {\epem} three jet events.
I fit an analytic expression for the multiplicity in three jet events
to event samples generated with QCD multihadronic
event generators.
I demonstrate that a one parameter fit of {$\cacf$}
yields the expected result {$\cacf$}=2.25 
in the limit of asymptotically large energies
if energy conservation is included in the calculation.
In contrast,
a two parameter fit of {$\cacf$} and a constant 
offset to the gluon jet multiplicity,
proposed in a recent study,
does not yield {$\cacf$}=2.25 in this limit.
I apply the one parameter fit method
to recently published data
of the DELPHI experiment at LEP and determine the effective
value of {$\cacf$} from this technique,
at the finite energy of the Z$^0$ boson,
to be $1.74\pm0.03\pm0.10$,
where the first uncertainty is statistical and the
second is systematic.

\end{titlepage}
 
\section{ Introduction }
\label{sec-intro}
 
At the basis of Quantum Chromodynamics (QCD),
the gauge theory of strong interactions,
are the color factors {$\caa$} and {$\cff$},
with values 3 and 4/3, respectively~\cite{bib-qcd}.
{$\caa$} determines the relative probability for a soft gluon
to couple to another gluon,
while {$\cff$} determines the corresponding probability
for a soft gluon to couple to a quark.
The ratio {$\cacf$} is perhaps the most fundamental quantity
in QCD besides the strong interaction coupling strength, $\alpha_S$.
Currently, the most accurate measurements of {$\cacf$} are
from angular correlations between jets in
four jet {\epem} events~\cite{lep-4jet}
and from the ratio of soft particle multiplicities 
at large transverse momenta to the jet axes between
unbiased gluon and quark jets~\cite{bib-opal99}.

Recently,
a new method to measure {$\cacf$} was
proposed~\cite{bib-delphi99},
based on the scale dependence of the mean 
particle multiplicity in {\epem} three jet events,
{$\nthreejet$}.
This method utilizes a next-to-leading-order 
(NLO\footnote{Also referred to as MLLA.})
analytic expression for {$\nthreejet$}~\cite{bib-khoze},
in conjunction with a constant offset term
{$\nzerog$}~\cite{bib-delphi99}
for the gluon jet multiplicity,
to perform a two parameter fit 
of {$\cacf$} and {$\nzerog$}.
The constant {$\nzerog$}
is intended to account for non-perturbative effects.
The variable {$\cacf$} is introduced using an analytic
expression for the mean multiplicity ratio 
between gluon and quark jets, $r$.
The expression used for $r$~\cite{bib-mueller}
does not incorporate recoil effects (energy conservation).

In this paper, I examine the determination
of {$\cacf$} from multiplicity in three jet events.
My principal purpose is to test the analytic expressions for 
{$\nthreejet$} and $r$.
The theoretical expressions are tested by fitting them
to event samples generated with QCD multihadronic
event generators.
The main conclusions are that 
to obtain the correct asymptotic result {$\cacf$}=9/4=2.25
from the method it is necessary to
use the pure QCD result~\cite{bib-khoze} without
the offset term~{$\nzerog$},
and to include recoil effects in the expression for $r$.
As a consistency check,
I apply my method to Monte Carlo events with
$\caa$=$\cff$=4/3 to verify that the fitted result
for the parameter $\cacf$ is consistent with unity
in this case.

Having established a fitting technique
that yields the correct results in the limiting cases of
(1)~asymptotically large energies, and
(2)~$\caa$=$\cff$,
I apply the method to recently published data~\cite{bib-delphi99}
of the DELPHI experiment at the LEP Collider at CERN.
I thereby determine the effective value of {$\cacf$}
from this method
at the finite energy of the Z$^0$ boson.

\section{Theoretical framework}
\label{sec-theory}

An analytic expression for the topology dependence of 
the mean particle multiplicity in {\epem} three jet 
quark-antiquark-gluon {\qqg} events, $\nthreejet$,
valid in the NLO approximation of perturbation theory,
is given by eq.~(6.43) of~\cite{bib-khoze}:
(see also~\cite{bib-dokshitzer}):
\begin{equation}
  \nthreejet = 
    \nee(\twoqscale) + r(\gscale)\, \frac{\nee(\gscale)}{2}
                \;\;\;\; ,
  \label{eq-threejet}
\end{equation}
where $\nee (Q)$ is the mean inclusive particle 
multiplicity of {\epem} annihilation events at energy scale $Q$.
The quark and gluon jet scales {$\qscale$} and 
{$\gscale$} are
(see eqs. (6.38) and (6.41) of~\cite{bib-khoze}):
\begin{eqnarray}
   \qscale & = & 
       \sqrt{
       \frac{ p_q \cdot p_{\overline{q}} }{ 2 } 
       } \;\;\;\; ;
           \label{eq-qscale} \\
   \gscale & = & 
       \sqrt{
       \frac{ 2\,(p_q\cdot p_g)\,(p_{\overline{q}}\cdot p_g) }
            { p_q \cdot p_{\overline{q}} } 
        }   \;\;\;\; , \label{eq-gscale}
\end{eqnarray}
with $p_q$, $p_{\overline{q}}$ and $p_g$
the 4-momenta of the $q$, ${\overline{q}}$ and $g$.
{$\qscale$} is the energy of the quark or antiquark
in the {\qq} rest frame,
while {$\gscale$} is the transverse momentum of the gluon
with respect to the {\qq} axis in that frame.
These equations are valid for massless quarks and gluons.

The quantity $r(Q)$ is the ratio of the mean multiplicities
between gluon and quark jets.
It has been calculated analytically in
the next-to-next-to-next-to-leading-order 
(3NLO) approximation of perturbation theory,
including recoil effects~\cite{bib-capella}:
\begin{equation}
    r(Q) = r_0\, (1 - r_1\gamma_0 
         - r_2\gamma_0^2 - r_3\gamma_0^3)
      \;\;\;\; ,
    \label{eq-ratio}
\end{equation}
where $\gamma_0(Q)$=$\sqrt{ 2\caa\alpha_S(Q)/\pi }$,
$r_0$={$\cacf$},
and the correction terms $r_1$, $r_2$ and $r_3$
are constants in QCD which functionally depend
on the color factors through terms proportional 
to $1/r_0$ and~$1/\caa$~\cite{bib-capella}.
$r$ depends on the scale $Q$ only through $\alpha_S$:
\begin{equation}
  \alpha_{S}(Q) = \frac {2\pi }{\beta_{0}y}
      \left[1-\frac{\beta_{1}\ln (2y)}
      {\beta_{0}^{2}y}\right] 
  \;\;\;\; ,  
  \label{eq-alphas}
\end{equation}
with $y=\ln(Q/\Lambda)$,
$\Lambda$ a cutoff which defines
the limit of perturbative evolution,
$\beta_{0}=(11\caa-2n_{f})/3$,
$\beta_{1}=[17(\caa)^2- n_{f}\caa(5+3/r_0)]/3$,
and n$_{\mathrm{f}}$ the number of active quark flavors.
In this paper, I set n$_{\mathrm{f}}$=5 and use the corresponding result
for $\Lambda$ found in a fit of the 3NLO expression for
quark jet multiplicity~\cite{bib-dgary99} to inclusive
{\epem} data.
This result,
$\Lambda$=0.148~GeV~\cite{bib-dgary99},
is similar to the value 
of~{$\lmsbar$}~\cite{bib-pdg}.\footnote{$\Lambda$ 
and $\lmsbar$ are strongly related to each other
but are not necessarily the same.}

\section{Analysis technique}
\label{sec-technique}

Three jet events are selected using standard jet finding algorithms
(see section~\ref{sec-selection}).
Two of the jets are identified as the quark
($q$ or $\overline{q}$) jets,
the other as the gluon jet.
The 4-momentum of each jet is assigned to the
underlying $q$, $\overline{q}$ or $g$.
Since expressions~(\ref{eq-qscale}) and~(\ref{eq-gscale})
are based on massless kinematics,
the jet momenta are modified to obtain massless jets.
First, the jets are assigned calculated energies, $E_{calc.}$,
based on the angles between jets,
assuming the jets are massless (see e.g.~\cite{bib-opal91}).
Second, the jet 3-momenta are scaled as follows:
\begin{equation}
  \vec{P} = \frac{E_{calc.}}{|\vec{P}_{jet-finder}|}\vec{P}_{jet-finder}
   \;\;\;\; ,
\end{equation}
with $\vec{P}_{jet-finder}$ the jet 3-momentum determined
by the jet finder.
The quark and gluon
4-momenta defined by $p$=($E_{calc.}$,$\vec{P}$)
are used to determine the 
scales~(\ref{eq-qscale}) and~(\ref{eq-gscale}).
This method of defining massless jets is referred to
in the literature as the E0 scheme~\cite{bib-ezero}.

The values of $\nee(\twoqscale)$ and $\nee(\gscale)$ 
in eq.~(\ref{eq-threejet})
are determined using
parametrizations of $\nee(Q)$ versus~$Q$.
These parametrizations are based on sixth order 
polynomials in $\ln (Q)$.
A parametrization is determined independently
for each Monte Carlo event sample\footnote{Herwig 
at the parton and hadron levels,
and Jetset at the parton level with $\caa$=$\cff$,
see section~\ref{sec-selection}.} 
and for the data.
For the Monte Carlo samples,
the parametrizations are obtained by a fit to the
predicted values of {$\nee$} versus 
$Q$={\ecm} in the interval
between 10~GeV and 10~TeV,
where {\ecm} is the center-of-mass (c.m.) energy.
For the data,
a fit is made to measurements of {$\nee$}
for 12~GeV$\,\leq\,${\ecm}$\,\leq\,$189~GeV.\footnote{The data used 
are the same as presented in Fig.~2 of~\cite{bib-dgary99}.}
The parametrizations provide good representations of
the multiplicity in all cases.

The analytic expression for $r$ (eq.~(\ref{eq-ratio}))
is introduced into eq.~(\ref{eq-threejet}).
$r_1$, $r_2$ and $r_3$ in expression~(\ref{eq-ratio})
depend on $1/r_0$ and~$1/\caa$,
as stated above.
Similarly,
$\beta_1$ in (\ref{eq-alphas}) depends on $1/r_0$,
while $\beta_0$ and $\beta_1$ depend on~{$\caa$}.
The leading $r_0$ term in (\ref{eq-ratio}) and the
$1/r_0$ terms in $r_1$, $r_2$, $r_3$ and $\beta_1$
form the fitted parameter.
$\caa$ in $r_1$, $r_2$, $r_3$, $\beta_0$ and $\beta_1$
is set equal to its QCD value of~3.
$r_0$ is then determined in a one parameter fit of
eq.~(\ref{eq-threejet}) to the Monte Carlo or experimental 
results for $\nthreejet$ as a function of scale.
The DELPHI Collaboration recently presented a
similar study~\cite{bib-delphi99}.
I discuss the DELPHI method and results 
in section~\ref{sec-delphi}.

\section{Monte Carlo samples and event selection}
\label{sec-selection}

For the principal Monte Carlo based results I present,
I use event samples generated with the
Herwig Monte Carlo multihadronic
event generator~\cite{bib-herwig}, version~5.9.
The parameter values used for Herwig are the same
as those given in~\cite{bib-opal99}.
Herwig contains the most complete computer simulation 
of QCD presently available,
including terms up to and beyond 
the next-to-next-to-leading order (NNLO) approximation.
In this sense Herwig resembles an analytic calculation.
In addition, Herwig implements exact energy-momentum 
conservation at each parton branching
and a model for hadronization.
Herwig yields the correct asymptotic result of 2.25 for
the multiplicity ratio $r$~\cite{bib-opal96}
and related quantities~\cite{bib-opal99}.
It provides a good description of gluon and quark jet
properties up to the highest available {\epem} 
energies.
Thus, Herwig provides a suitable QCD reference sample.
Herwig generally predicts that QCD variables reach their
asymptotic values at center of mass energies
of several TeV or more,
depending on the variable.
In the following,
I choose 10~TeV as the canonical c.m. energy at which to
test my fit method in the asymptotic limit.

For studies with $\caa$=$\cff$=4/3,
I employ a special version of the
Jetset Monte Carlo muiltihadronic
event generator~\cite{bib-jetset},
version~7.4, with parameter values given in~\cite{bib-opal96b}.
In addition to setting $\caa$=$\cff$,
I turn off gluon splittings, $g\rightarrow\,${\qq}.
The reason Jetset is used for these studies,
and not Herwig,
is that Herwig does not allow $\caa$=$\cff$.
Jetset is based on leading order (LO) QCD
with a simulation of coherence effects due to
higher orders.
The standard version of Jetset does not yield 
the correct asymptotic result for $r$,
as seen from Fig.~2 of~\cite{bib-opal96},
except perhaps at exceptionally high energies
({\ecm}$>>$100~TeV~??).
Thus the QCD predictions of Jetset 
should be treated with precaution.
For the present purposes
it is sufficient that quark and gluon jets have 
the same internal properties,
such as multiplicity,
if $\caa$=$\cff$.
This is satisfied by the special
version of Jetset at the parton level.
By parton level,
I mean the ensemble of quarks and gluons
which are present at the termination of
the perturbative stage of evolution.

\begin{figure}[p]
\begin{center}
  \epsfxsize=23cm
  \epsffile[100 0 630 550]{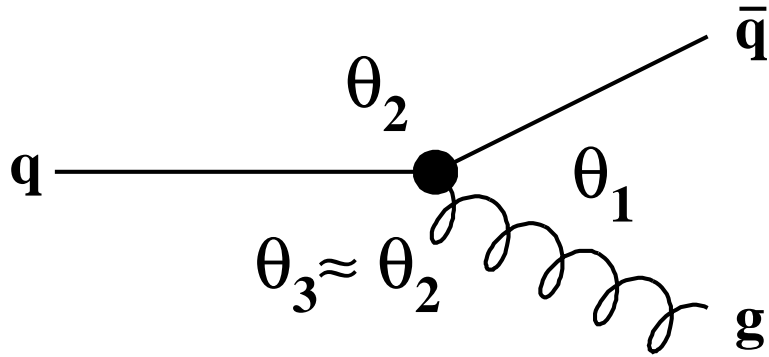}
\end{center}
\vspace*{-5in}
\caption{
Schematic representation of a three jet {\qqg} event
produced in {\epem} annihilations with a 
Y event topology~\cite{bib-opal91},
in which the angle between the highest energy jet
and each of the two lower energy jets is about the same.
The angle $\theta_1$ opposite the highest energy jet
is used to specify the event topology.
\label{fig-yevent}
} 
\end{figure}

Three jet events are constructed from these samples
by adjusting the resolution scale(s) of a jet finding algorithm
for each event so that exactly three jets are found.
I choose three jet finding algorithms:
the {\durham}~\cite{bib-durham}, Jade~\cite{bib-jade} 
and cone~\cite{bib-cone} jet finders.
These three algorithms are very different in their
treatment of soft particles.
The difference in the results found using the three algorithms
therefore provides a rigorous test of
the jet finder independence of the method.
I retain events if 
the angle between the highest energy jet and each of
the two other jets is the same to within~5$^\circ$,
the so-called ``Y events.''\footnote{Y events
were first studied in~\cite{bib-opal91}.}
An example of a Y event is shown in Fig.~\ref{fig-yevent}.
Measurements of the particle multiplicity of Y events
as a function of topology (i.e.~scale)
have recently become available~\cite{bib-delphi99}.
I wish to utilize these data for my fits 
(section~\ref{sec-dataresults}).
This provides my principal motivation for selecting Y events.
For Y events,
the three jet event multiplicity $\nthreejet$ and the scales
(\ref{eq-qscale}) and~(\ref{eq-gscale}) depend only on
{\ecm} and one angle in the event,
conveniently chosen to be~$\theta_1$
(see Fig.~\ref{fig-yevent}).
For fixed {\ecm},
$\theta_1$ therefore determines the scale.

For the Monte Carlo events used here,
the quark and gluon jets are identified using 
parton level Monte Carlo information.
The directions of the primary quark and 
antiquark\footnote{i.e., the $q$ and $\overline{q}$
produced directly in the electroweak decay of the
virtual Z$^0/\gamma$ in 
{\epem}$\,\rightarrow\,$(Z$^0/\gamma$)$^*$$\,\rightarrow\,hadrons$ events.}
are determined
after their perturbative evolution has terminated.
The jet closest to the direction of the evolved primary
quark or antiquark is considered to be a quark jet.
The distinct jet closest to the other evolved primary
quark or antiquark is considered to be the other quark jet.
The remaining jet is identified as the gluon jet.
This algorithm is applied to jets at both 
the parton and hadron levels.
By hadron level,
I mean the level after hadronization,
with charged and neutral particles with lifetimes greater than
3$\,\times\,$10$^{-10}$~s treated as stable.
Hence charged particles from the decays of K$^0_{\mathrm{s}}$
and weakly decaying hyperons are included in the definition 
of the hadron level multiplicity.


\section{Monte Carlo based results}
\label{sec-mcresults}

\begin{figure}[p]
\begin{center}
  \epsfxsize=15cm
  \epsffile[45 120 595 720]{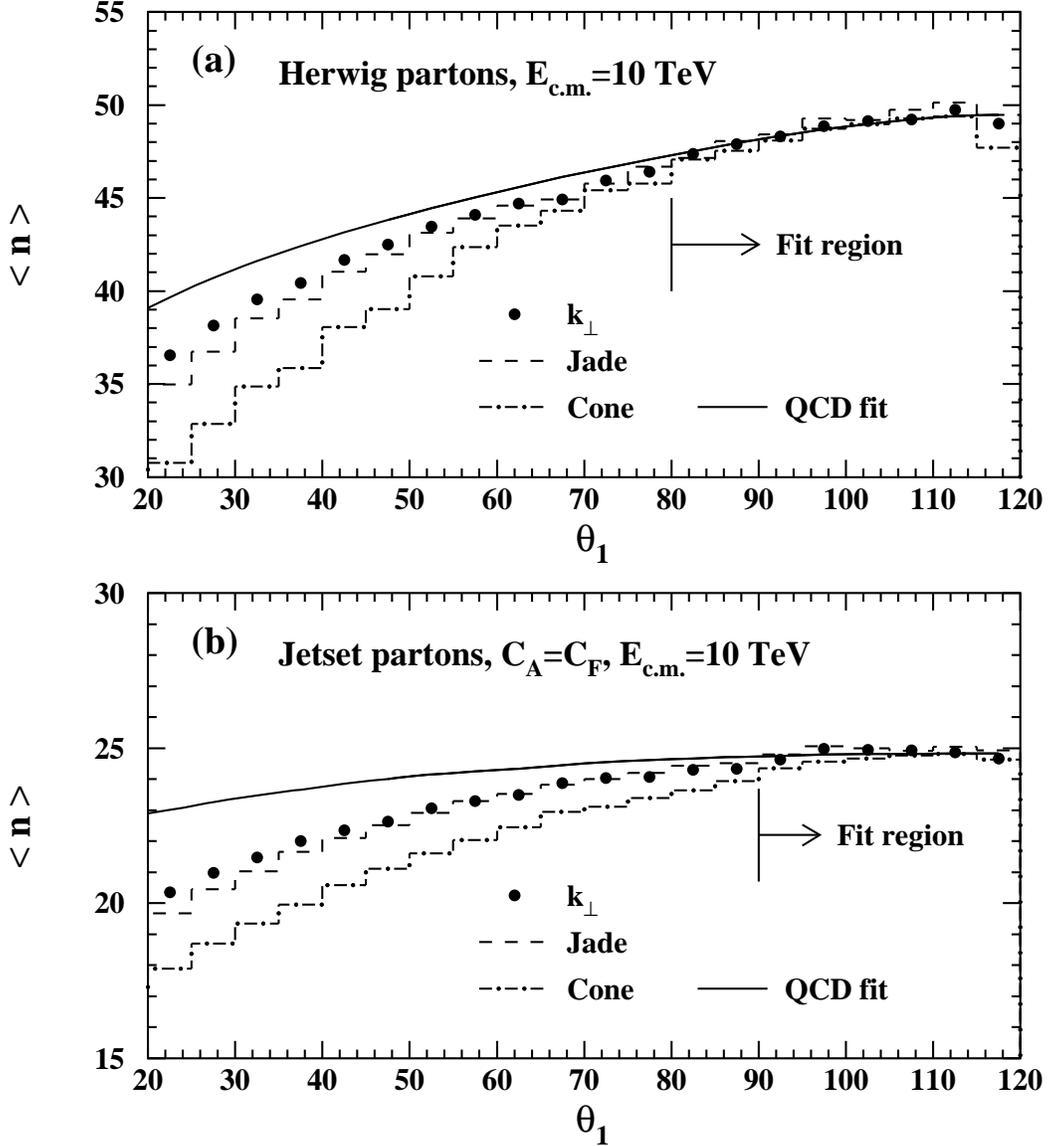}
\end{center}
\caption{
(a) The mean parton level multiplicity 
of three jet Y events as a function
of the opening angle $\theta_1$,
for events generated with the
Herwig multihadronic event generator.
(b)~The analogous results for events generated with the
Jetset multihadronic event generator with $\caa$=$\cff$.
The event samples in (a) and (b) are selected using
the {\durham}, Jade and cone jet finders.
The c.m. energy is 10~TeV.
The solid curves show the results of a one parameter QCD fit 
to events selected using the {\durham} jet finder.
The fits are performed within the regions shown.
\label{fig-10tev}
} 
\end{figure}

I begin by studying Herwig events at the parton level,
with {\ecm}=10~TeV.
This large energy is meant to ensure that the fit
results are asymptotic, as mentioned above.
The mean multiplicity of these events as a function of the 
opening angle~$\theta_1$ is shown in Fig.~\ref{fig-10tev}a.
The results are shown for the three jet algorithms.
The results of the three algorithms are seen to be similar
for angles larger than about 80$^{\circ}$.
As $\theta_1$ becomes smaller,
the two lower energy jets are not as well separated
and background from two jet-like events increases.
Different jet finders have different efficiencies 
for selecting background:
thus the results of the jet finders diverge.
Since the results should be independent of the choice
of a jet algorithm to be sensible,
I restrict the fits to the region where the predictions of the
jet finders approximately agree,
namely $80^\circ$$\,\leq\,$$\theta_1$$\,\leq\,$$120^\circ$.

The solid curve in Fig.~\ref{fig-10tev}a shows the
result of the one parameter fit of eq.~(\ref{eq-threejet})
to the event multiplicity determined using the {\durham} jet finder.
The curve provides a reasonable description of
the multiplicity inside the fit region.
Outside this region,
i.e.~for angles less than $80^\circ$,
the fitted curve does not describe the event multiplicity well.
The multiplicity of the events in Fig.~\ref{fig-10tev}a
is not well defined for $\theta_1 <80^\circ$, however,
since the results from different jet finders disagree strongly.
Therefore I do not consider the discrepancies between the
fitted curve and the jet finder based results
for $\theta_1 <80^\circ$ to be meaningful.

The results for $r_0$
are summarized in the top portion of Table~\ref{tab-r0values}.
Taking the result found using the {\durham} jet finder
as the central value,
with half the difference between the extreme values found
using the different jet finders as a systematic uncertainty,
I obtain $r_0$=$2.248\pm0.010\,$(stat.)$\pm0.024\,$(syst.),
consistent with the QCD asymptotic expectation of~2.25.

The analogous results for Jetset at the parton level
with $\caa$=$\cff$ are shown in Fig.~\ref{fig-10tev}b.
Again, the c.m. energy is 10~TeV.
The predictions of the three jet finders are seen to be similar
only for $\theta_1>90^\circ$.
Therefore, I limit the fit range to
$90^\circ$$\,\leq\,$$\theta_1$$\,\leq\,$$120^\circ$ in this case.
The results for $r_0$ are given in the bottom portion
of Table~\ref{tab-r0values}.
Combining the results in the manner described
in the previous paragraph yields
$r_0$=$1.012\pm0.009\,$(stat.)$\pm0.027\,$(syst.),
consistent with unity.

Thus a one parameter fit of $r_0$
yields the correct results in the limiting cases of
(1)~QCD at asymptotically large energies and (2)~$\caa$=$\cff$,
as long as the fit range is restricted to regions where the
results of the different jet finders agree, or,
equivalently,
to regions where the fitted curves
provide a good description of the multiplicity.
The fits generally yield
$\chi^2/d.o.f.$$\,\sim\,$1 (Table~\ref{tab-r0values}).

It is of interest to determine the importance of 
energy conservation in the expression for~$r$.
To this effect,
I replace eq.~(\ref{eq-ratio}) by the corresponding result 
in the NNLO approximation both with~\cite{bib-dn} and
without~\cite{bib-mueller} recoil effects,
and repeat the fit of Herwig events described above.
The NNLO approximation is used for this test,
and not the 3NLO approximation,
because a 3NLO expression for $r$ 
without energy conservation is not available.
For the {\durham} based event sample,
the NNLO calculation with energy conservation yields 
$r_0$=$2.254\pm0.010\,$(stat.),
not very different from the 3NLO result presented above.
The corresponding result without energy conservation is
only $2.079\pm0.009\,$(stat.), however,
significantly smaller than~2.25.
This implies that it is important to include energy conservation
in the analytic expressions,
even for {\ecm}=10~TeV.

\begin{figure}[p]
\begin{center}
  \epsfxsize=15cm
  \epsffile[45 120 595 720]{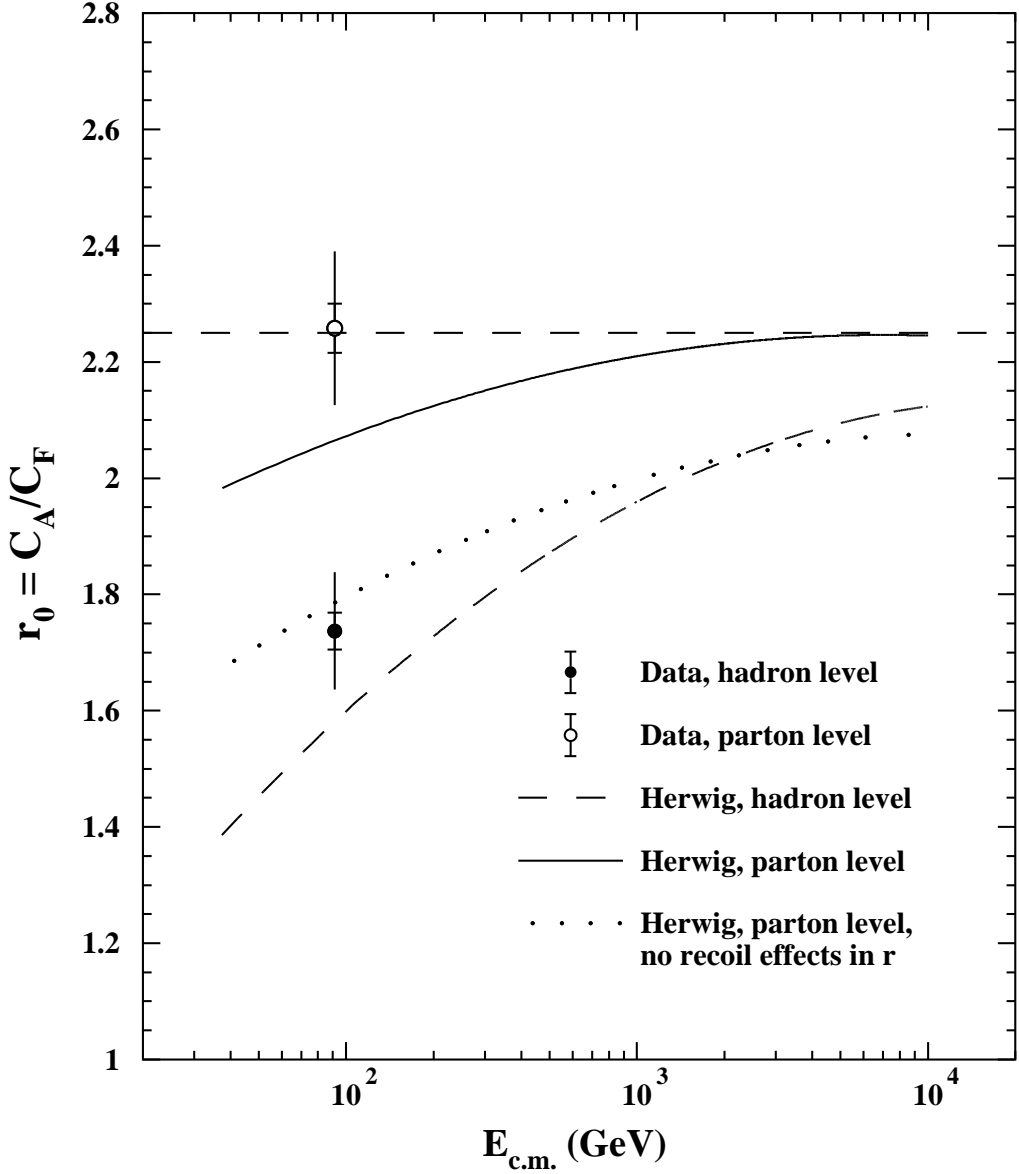}
\end{center}
\caption{Results of a one parameter fit of $r_0$={$\cacf$} 
as a function of the c.m. energy,
for Herwig Monte Carlo three jet Y
events at the parton and hadron levels.
The events are selected using the {\durham} jet finder.
The corresponding results for data~\cite{bib-delphi99}
at {\ecm}=91~GeV
are shown by the open and solid points.
The hadron level results are based on charged particles only.
For the data points, the vertical lines show the total uncertainties,
with statistical and systematic terms added in quadrature.
The small horizontal lines indicate the statistical uncertainties.
\label{fig-ecm}
} 
\end{figure}

In Fig.~\ref{fig-ecm},
I show the fitted results for $r_0$ as a function of~{\ecm},
using Herwig events at the parton and hadron levels.
The events are selected using the {\durham} jet finder.
The hadron level multiplicity is based on charged particles only.
The fit interval is
$80^\circ$$\,\leq\,$$\theta_1$$\,\leq\,$$120^\circ$,
i.e.~the same as in Fig.~\ref{fig-10tev}a.
The fitted curves provide
good descriptions of the multiplicity within the fit region
for all c.m. energies at both the parton and hadron levels.
The fit results are observed to have only
a moderate dependence on the choice of the jet algorithm,
generally similar to that indicated in Table~\ref{tab-r0values}
for parton level events
or in item~1 of the list presented below
in section~\ref{sec-dataresults} for hadron level events.
From the parton level curve (solid line) it is seen that
the asymptotic result $r_0$$\,\approx\,$2.25
is reached for {\ecm}$\,\sim\,$3~TeV.
The hadron level curve (dashed line)
converges to the asymptotic limit much more slowly, however.
As a consequence,
the hadronization correction,
defined by the ratio of the parton 
to the hadron level results,
is fairly large.
The hadronization correction is predicted to be
1.30 at the mass of the Z$^0$ and 1.06 at 10~TeV.

The dotted curve in Fig.~\ref{fig-ecm} shows the fitted results
for $r_0$ at the parton level if
the NNLO expression for $r$ {\it without} recoil effects
is used in place of the 3NLO expression.
The QCD asymptotic limit of 2.25 it not attained in this case,
again emphasizing the importance of energy conservation.

\section{Data based results}
\label{sec-dataresults}

Recently,
the DELPHI experiment at LEP presented measurements
of the charged particle multiplicity of Y events
and the scales~(\ref{eq-qscale}) and~(\ref{eq-gscale})
as a function of the opening angle~$\theta_1$~\cite{bib-delphi99}.
The results are based on the {\durham} jet finder 
with {\ecm}=91~GeV.
These data allow me the possibility
to determine the effective value of $r_0$ at the scale 
of the~Z$^0$ using my one parameter fit method.
The DELPHI multiplicity measurements
are shown in Fig.~\ref{fig-delphi}.
The result of the one parameter fit
is shown by the solid curve.
The fit range employed is 
$78^\circ$$\,\leq\,$$\theta_1$$\,\leq\,$$120^\circ$,
similar to the interval of 
$80^\circ$$\,\leq\,$$\theta_1$$\,\leq\,$$120^\circ$
I choose for Herwig events
(Figs.~\ref{fig-10tev}a and~\ref{fig-ecm}).
The small difference in the choice of fit interval
between the Herwig and DELPHI samples is
not important (see item~2 below).
The analytic curve provides a good description 
of the measurements within the fit region,
yielding $\chi^2/d.o.f.$=8.9/8.

The result for the fitted parameter is
$r_0$=$1.737\pm0.032\,$(stat.).
To estimate a systematic uncertainty for this result,
I consider the following.
\begin{enumerate}
  \item {\it Jet finder dependence.}
    The DELPHI results are presented for the {\durham} jet
    finder, but not for the Jade or cone jet finders.
    Herwig at the hadron level with {\ecm}=91~GeV yields
    $r_0$=1.585 for the {\durham} jet finder,
    1.601 for the Jade jet finder,
    and 1.516 for the cone jet finder,
    where the statistical uncertainty is 0.008 in all cases.
    Half the difference between the extreme values is
    taken as a systematic uncertainty.
  \item {\it Fit interval.}
    The fit interval I choose for the DELPHI data 
    is $78^\circ$$\,\leq\,$$\theta_1$$\,\leq\,$$120^\circ$, as stated above.
    Decreasing the lower limit of this
    interval to 60$^\circ$ yields $r_0$=$1.705\pm0.025\,$(stat.),
    while decreasing the upper limit to 90$^\circ$,
    with the lower limit at the standard value,
    yields $r_0$=$1.755\pm0.038\,$(stat.).
    I take half the difference between these values
    as a systematic uncertainty.
    A further test of the choice of the fit interval is presented in
    Fig.~\ref{fig-fitrange}.
    Fig.~\ref{fig-fitrange}a shows the fitted 
    results for $r_0$ as a function of $\theta_1^{\mathrm{min.}}$,
    where $\theta_1^{\mathrm{min.}}$ is the lower limit of the fit range
    $\theta_1^{\mathrm{min.}}$$\,\leq\,$$\theta_1$$\,\leq\,$$120^\circ$.
    The corresponding values of $\chi^2/d.o.f.$ are shown
    in Fig.~\ref{fig-fitrange}b.
    The $\chi^2/d.o.f.$ is 1.4 or less 
    for $\theta_1^{\mathrm{min.}}\geq 60^\circ$ but 
    much larger for $\theta_1^{\mathrm{min.}}<60^\circ$.
    This provides justification for not extending
    the fit range below~$60^\circ$,
    i.e.~the fit is restricted to an interval where the theoretical
    expression eq.~(\ref{eq-threejet}) describes the data accurately.
  \item {\it Parametrization of $\nee$ versus $Q$.}
    Rather than use a polynomial parametrization of $\nee$ versus $Q$
    (section~\ref{sec-technique}),
    I use the parametrization based on the 3NLO expression
    for quark jet multiplicity~\cite{bib-dgary99}
    with the parameter values in~\cite{bib-dgary99}.
    This yields $r_0$=$1.804\pm0.032\,$(stat.).
    The difference with respect to the standard result is
    taken as a systematic uncertainty.
    Note that the polynomial provides a better
    description of $\nee$ versus $Q$ than the 3NLO expression.
  \item {\it Value of $\Lambda$.}
    Setting $\Lambda$ in eq.~(\ref{eq-alphas}) to the PDG
    value of {$\lmsbar$}=0.220~GeV~\cite{bib-pdg},
    rather than using  0.148~GeV (section~\ref{sec-theory}),
    yields $r_0$=$1.761\pm0.033\,$(stat.).
    I take the difference with respect to the standard result
    as a systematic uncertainty.
  \item {\it Averaging procedure for $\qscale$ and $\gscale$.}
    The DELPHI results 
    for the quark and gluon scales
    eqs.~(\ref{eq-qscale}) and~(\ref{eq-gscale})
    are found by taking the {\it geometric means} of
    $\qscale$ and $\gscale$, averaged over the event sample,
    as a function of~$\theta_1$.
    For the Monte Carlo based results 
    presented in section~\ref{sec-mcresults},
    I employ the much more common {\it arithmetic means}.
    For hadron level events at 91~GeV,
    Herwig with the {\durham} jet finder yields
    $r_0$=1.629 for geometric means and
    $r_0$=1.585 for arithmetic means,
    where the statistical uncertainty is~0.008 in both cases.
    The difference between these values
    is taken as a systematic uncertainty.
  \item {\it Number of active flavors,} n$_{\mathrm{f}}$.
    Using n$_{\mathrm{f}}$=3 rather than 
    n$_{\mathrm{f}}$=5 in the analytic expressions
    for $r$ and $\alpha_S$ (eqs.~(\ref{eq-ratio}) and~(\ref{eq-alphas})),
    and correspondingly evaluating~$\alpha_S$
    using $\Lambda$=0.322 GeV \cite{bib-dgary99}
    rather than $\Lambda$=0.148~GeV,
    yields $r_0$=$1.735\pm0.029\,$(stat.).
    The difference with respect to the standard result is taken
    as a systematic uncertainty.
\end{enumerate}
The systematic uncertainties are summarized in Table~\ref{tab-systerr}.
The largest terms arise from the parametrization of $\nee$,
the averaging procedure for the scales,
and the choice of jet finder, in that order.
The terms are added in quadrature to define the
total systematic uncertainty.
The final result for the effective value of $r_0$=$\cacf$
at the scale of the Z$^0$ is:
\begin{equation}
   r_0 = 1.737\pm0.032\,{\mathrm{(stat.)}}\pm0.097\,{\mathrm{(syst.)}}
    \;\;\;\; .
\end{equation}
Multiplying this value by the hadronization correction of 1.30 
mentioned at the end of section~\ref{sec-mcresults} yields
$r_0 = 2.26\pm0.04\,{\mathrm{(stat.)}}\pm0.12\,{\mathrm{(syst.)}}$
as the corresponding result at the parton level,
consistent with the QCD asymptotic prediction of~2.25.
The data based results I obtain
at the hadron and parton levels 
are shown by the solid and open points in
Fig.~\ref{fig-ecm}.
The experimental results lie somewhat above the
Herwig curves but are generally consistent with them.

\section{Comparison to a two parameter fit method}
\label{sec-delphi}

\begin{figure}[p]
\begin{center}
  \epsfxsize=15cm
  \epsffile[45 120 595 720]{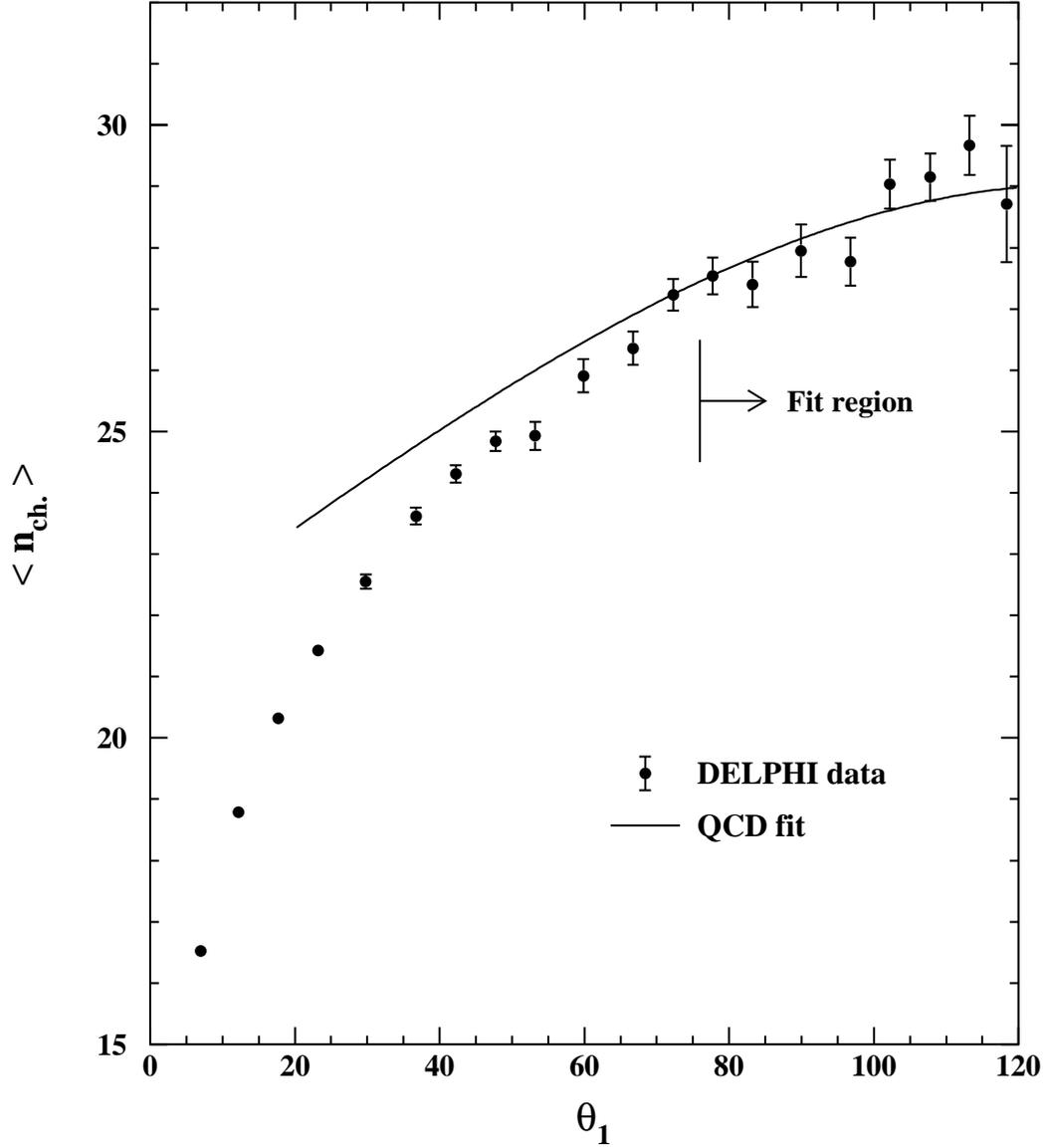}
\end{center}
\caption{
Measurements~\cite{bib-delphi99} of the
mean charged particle
multiplicity of three jet Y events as a function of the
opening angle $\theta_1$,
for {\ecm}=91~GeV.
The events are selected using the {\durham} jet finder.
The solid curve shows the result of a one parameter
fit to the data within the fit region shown.
\label{fig-delphi}
} 
\end{figure}

\begin{figure}[p]
\begin{center}
  \epsfxsize=15cm
  \epsffile[45 120 595 720]{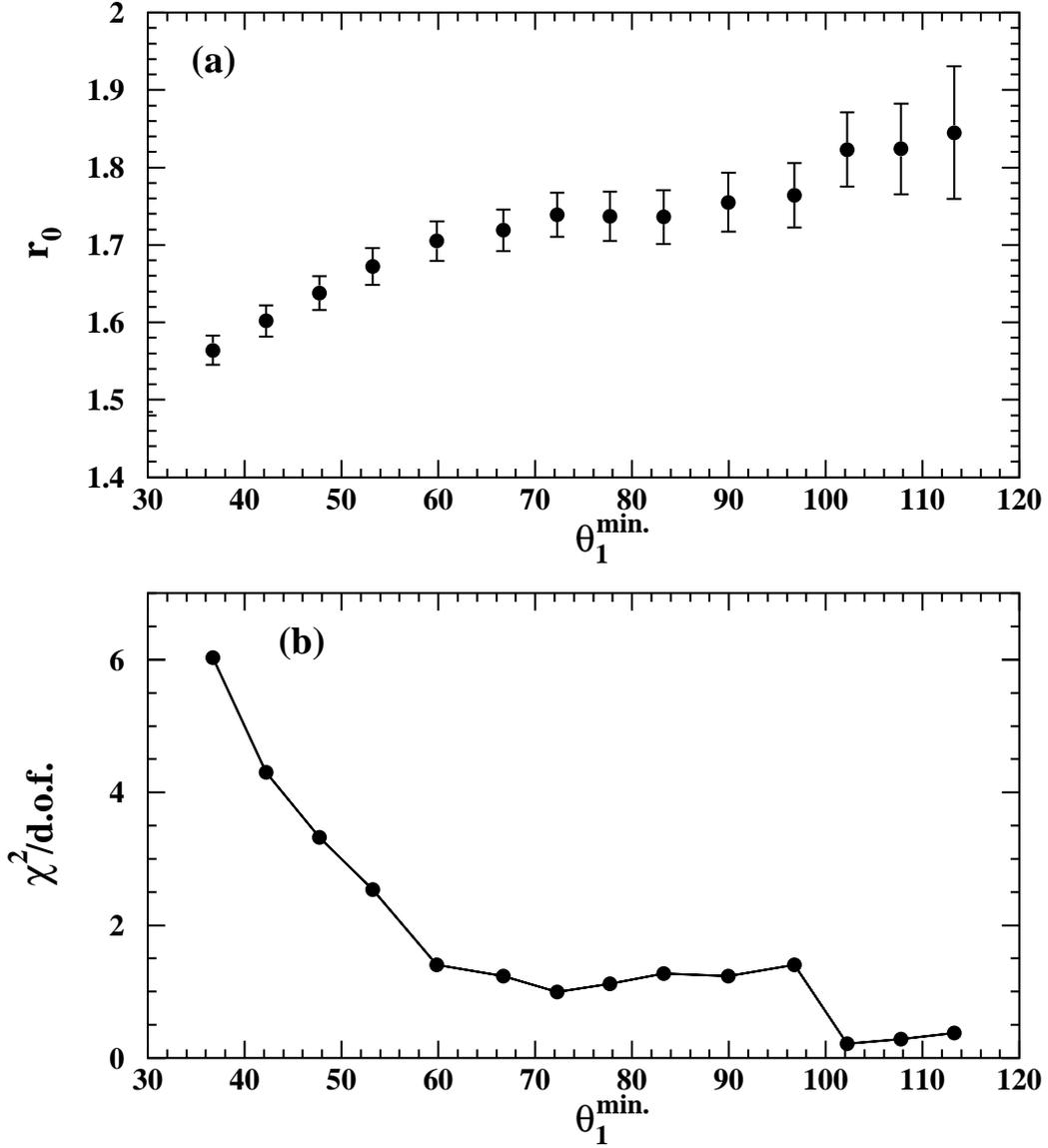}
\end{center}
\caption{
(a) Results of a one parameter fit of $r_0$={$\cacf$} to
measurements~\cite{bib-delphi99}
of the charged particle multiplicity of three jet
Y events at {\ecm}=91~GeV,
as a function of the lower limit $\theta_1^{\mathrm{min.}}$
of the fit range
$\theta_1^{\mathrm{min.}}$$\,\leq\,$$\theta_1$$\,\leq\,$$120^\circ$.
(b)~The corresonding values of $\chi^2/d.o.f$.
The events are selected using the {\durham} jet finder.
\label{fig-fitrange}
} 
\end{figure}

In their recent publication~\cite{bib-delphi99},
the DELPHI Collaboration presented an alternative 
method to determine {$r_0$=$\cacf$} 
from three jet event particle multiplicity.
I used the data of that study,
shown in Fig.~\ref{fig-delphi},
to obtain the results of section~\ref{sec-dataresults}.
The DELPHI analysis is based on a fit of the expression
\begin{equation}
  \nthreejet = 
    \nee(\twoqscale) + r(\gscale)\, 
         \left[ \frac{\nee(\gscale)}{2}-\nzerog\right]
  \label{eq-delphithreejet}
\end{equation}
to the three jet event multiplicity data,
where $\nzerog$ is a parameter meant to account for differences
in the hadronization of gluons and quarks.
The DELPHI analysis differs from mine principally by using
expression~(\ref{eq-delphithreejet})
rather than expression~(\ref{eq-threejet}),
by using the NNLO result for $r$
without energy conservation~\cite{bib-mueller}
rather than the 3NLO expression,
and by invoking a two parameter fit of $r_0$ and $\nzerog$
rather than a one parameter fit.
The DELPHI analysis also differs from mine in the choice of fit range
and in the parametrization of {$\nee$} versus~$Q$.
In the DELPHI study,
the fit range is
$30^\circ$$\,\leq\,$$\theta_1$$\,\leq\,$$120^\circ$
rather than $78^\circ$$\,\leq\,$$\theta_1$$\,\leq\,$$120^\circ$
and the parametrization of {$\nee$} versus~$Q$ is based on
the NLO expression for quark jet multiplicity
in {\epem} annihilations~\cite{bib-nlo} rather than a polynomial.
The DELPHI results utilize events at {\ecm}=91~GeV 
selected using the {\durham} jet finder,
as stated in section~\ref{sec-dataresults}.

\begin{figure}[p]
\begin{center}
  \epsfxsize=15cm
  \epsffile[45 120 595 720]{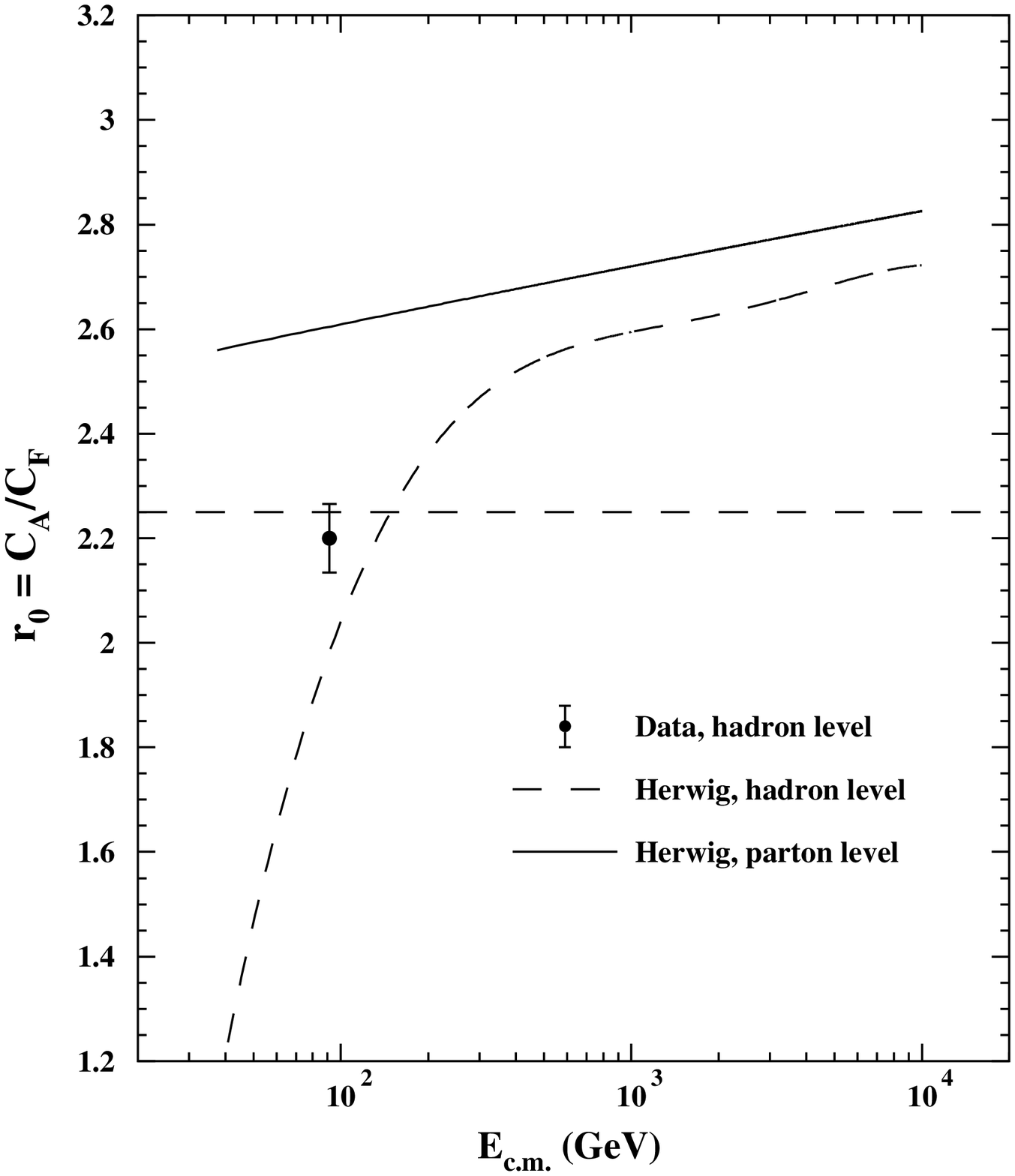}
\end{center}
\caption{
Results for $r_0$=$\cacf$ from a two parameter fit
method~\cite{bib-delphi99} as a function of the c.m. energy,
for Herwig Monte Carlo three jet Y events 
at the parton and hadron levels.
The events are selected using the {\durham} jet finder.
The corresponding hadron level
result found using data~\cite{bib-delphi99}
at {\ecm}=91~GeV is shown by the solid point.
The hadron level results are based on charged particles only.
The uncertainty shown for the data point is statistical.
\label{fig-delphir0}
} 
\end{figure}

Repeating the DELPHI analysis,
viz. a two parameter fit of eq.~(\ref{eq-delphithreejet})
to the data in Fig.~\ref{fig-delphi},
using the DELPHI values of {$\qscale$} and {$\gscale$},
the expression for $r$ in~\cite{bib-mueller},
a fit range from $30^\circ$ to~$120^\circ$,
and the NLO expression for quark jet multiplicity
to parametrize {$\nee$} versus~$Q$,\footnote{For the NLO 
parametrization of quark jet multiplicity,
I use the parameters in~\cite{bib-opal97}.}
I obtain
\begin{eqnarray}
   r_0 & = & 2.200\pm0.066\,(\mathrm{stat.})
     \;\;\;\; ;  \label{eq-delphir0} \\
   \nzerog & = & 1.46\pm0.10\,(\mathrm{stat.})
     \;\;\;\; .  \label{eq-delphin0g}
\end{eqnarray}
The $\chi^2/d.o.f.$ of the fit is 13.2/16.
The results (\ref{eq-delphir0}) 
and (\ref{eq-delphin0g}) are very similar
to those of DELPHI,
namely $r_0$=$2.251\pm0.063\,(\mathrm{stat.})$
and $1.40\pm0.10\,(\mathrm{stat.})$~\cite{bib-delphi99}.
The result (\ref{eq-delphir0}) for $r_0$ is shown 
by the solid point in Fig.~\ref{fig-delphir0}.

The value of $r_0$ derived from the DELPHI
two parameter fit method is numerically 
very similar to the QCD asymptotic result {$\cacf$}=2.25.
On this basis, DELPHI suggests~\cite{bib-delphi99}
that their analysis provides a measurement of that quantity.
To test this hypothesis,
I determine the results of the DELPHI method
in the two limiting cases discussed in section~\ref{sec-mcresults}:
(1)~asymptotically large energies and (2)~$\caa$=$\cff$.
For Herwig events at the parton level with {\ecm}=10~TeV,
the DELPHI fit method yields $r_0$=$2.80\pm0.03\,$,
3.19$\pm0.03\,$ and $4.03\pm0.05\,$
for events selected using the {\durham},
Jade and cone jet finders, respectively,
where the uncertainties are statistical.
These values are much larger than 2.25 and
exhibit a strong dependence on the choice of the jet finder,
in contrast to the results of
section~\ref{sec-mcresults}
(top portion of Table~\ref{tab-r0values}).
The analogous results for Jetset at the parton level
with {\ecm}=10~TeV and $\caa$=$\cff$ are $1.76\pm0.03\,$(stat.),
$2.05\pm0.02\,$(stat.), and $2.68\pm0.03\,$(stat.),
which are inconsistent with unity and
again exhibit a strong jet finder dependence.
This is also in contrast to the results of section~\ref{sec-mcresults}
(bottom portion of Table~\ref{tab-r0values}).
For the above results,
the NLO expression for quark jet multiplicity is fitted to
the MC predictions of $\nee$ versus {\ecm}
for both the Herwig and Jetset samples,
using scale values between 20~GeV and 10~TeV.
The results are similar if the polynomial parametrizations
discussed in section~\ref{sec-technique} are used instead.

The dashed and solid curves in Fig.~\ref{fig-delphir0} 
show the results I obtain for $r_0$ from applying the
DELPHI fit method to Herwig events at the hadron and parton levels.
The results are shown as a function of {\ecm}.
The hadron level results are based on charged particles only.
The event samples are selected using the {\durham} jet finder.
Thus Fig.~\ref{fig-delphir0} is the analogue for the DELPHI method
of the results I show in Fig.~\ref{fig-ecm} for my method.
The hadron level curve in Fig.~\ref{fig-delphir0} 
is seen to be in general agreement 
with the experimental result
(\ref{eq-delphir0}) at the scale of the~Z$^0$.
Asymptotically,
the hadron level prediction reaches a value
of about 2.7, however,
much larger than~2.25.
The parton level curve exceeds 2.25 by a 
large margin even at {\ecm}=91~GeV.

On the basis of the above results,
I conclude that the DELPHI fit method probably
does not measure $\cacf$,
and that the similarity of the hadron level result~(\ref{eq-delphir0})
to the asymptotic prediction
{$\cacf$}=2.25 is most likely a coincidence.
As a last note I remark that if energy conservation is
included in the NNLO expression for~$r$,
the result (\ref{eq-delphir0}) increases to
$r_0$=$2.479\pm0.081\,$(stat.).
Thus if energy conservation is incorporated 
into the DELPHI fit method,
the value of $r_0$ derived from charged hadrons at 91~GeV
is no longer similar to~2.25.

\section{Summary and Conclusions}
\label{sec-summary}

In this paper,
I have presented a test of the QCD expression for
the topology (scale) dependence of
particle multiplicity in {\epem} three jet events.
Using event samples generated with the Herwig Monte Carlo
multihadronic event generator as a reference,
I find that the QCD expression yields the correct
result {$\cacf$}=2.25 in the asymptotic limit of large energy
scales $Q$$\,\sim\,$3~TeV as long as it is used in
conjunction with an expression for $r$ which incorporates
energy conservation,
where $r$ is the ratio of mean particle multiplicities
between gluon and quark jets.
This emphasizes the importance of
energy conservation in QCD analytic expressions,
even at large scales.
My analysis is based on a one parameter fit
of {$\cacf$} to three jet event mean particle multiplicity
as a function of the topology of the event.

As a second test,
I apply my method to a sample of Monte Carlo
three jet events in which the color factors are set equal,
$\caa$=$\cff$.
I obtain {$\cacf$}$\,\approx\,$1 in this case,
demonstrating the self consistency of the technique.

Applying my fit method to recently 
published data~\cite{bib-delphi99}
of the DELPHI experiment at LEP,
I obtain
{$\cacf$}=$1.737\pm0.032\,{\mathrm{(stat.)}}\pm0.097\,{\mathrm{(syst.)}}$
as the effective value of the color factor ratio 
at {\ecm}=91~GeV from this technique.
This result is based on charged particles.
It is of interest to compare this result to related measurements
at {\ecm}=91~GeV
based on the charged particle multiplicity ratio
between gluon and quark jets,~$\rch$.
The experimental result for $\rch$ in full phase space is
$1.514\pm0.019\,{\mathrm{(stat.)}}
\pm0.034\,{\mathrm{(syst.)}}$~\cite{bib-opal99,bib-opal96}.
The corresponding result for $\rch$ in limited phase space,
defined by soft particles with large
transverse momenta to the jet axes,
is $2.29\pm0.09\,{\mathrm{(stat.)}}\pm0.15\,
{\mathrm{(syst.)}}$~\cite{bib-opal99}.
All these measurements 
--~the one presented here and the two based on~$\rch$~--
are predicted to 
equal 2.25 in the limit of large energies.
The result presented here is seen to be
intermediate to the two based on~$\rch$,
both in value and in the size of the uncertainty.
Whereas $\rch$ in limited phase space 
has already attained its asymptotic value at {\ecm}=91~GeV,
$\rch$ in full phase space and the result presented here
are sub-asymptotic at this scale.

Last, I test a two parameter fit method to determine
$\cacf$ from particle multiplicity in three jet events,
proposed in a recent study~\cite{bib-delphi99}.
I find that this method does not yield the correct results
{$\cacf$}$\,\approx\,$2.25 or {$\cacf$}$\,\approx\,$1
in the two limiting cases of QCD at asymptotic energies
or identical color factors $\caa$=$\cff$,
in contrast to my method.
Thus, I conclude that this two parameter fit method
probably does not measure the color factor ratio.

\section{Acknowledgements}

This work has been supported by
the US Department of Energy under grant DE-FG03-94ER40837.
I thank Marina Giunta for assistance in
the early stages of this analysis.


\pagebreak

\begin{table}[h]
\centering
\begin{tabular}{|l|cc|}
 \hline
  & & \\[-2.4mm]
  & $r_0$={$\cacf$} & $\chi^2/d.o.f.$ \\[2mm]
 \hline
 \hline
  Herwig partons & & \\
  {\durham} jet finder & $2.248\pm0.010$ &  5.9/8 \\
  Jade jet finder  & $2.269\pm0.010$ &  12.8/8 \\
  Cone jet finder  & $2.221\pm0.013$ & 14.7/8 \\
 \hline
  Jetset partons, {$\caa$}={$\cff$}& & \\
  {\durham} jet finder & $1.012\pm0.009$ &  5.9/6 \\
  Jade jet finder  & $1.032\pm0.007$ &  3.9/6 \\
  Cone jet finder  & $0.979\pm0.008$ &  6.7/6 \\
 \hline
\end{tabular}
\caption{Results of a one parameter fit of~$r_0$={$\cacf$} 
to the parton level multiplicity in three jet events,
as predicted by the Herwig QCD multihadronic event generator,
and by the Jetset multihadronic event generator with {$\caa$}={$\cff$}.
The {\ecm} value is 10~TeV for both samples.
The fits are performed using the 3NLO expression for the multiplicity
ratio between gluon and quark jets,~$r$.
The uncertainties are statistical.
}
\label{tab-r0values}
\end{table}

\begin{table}[h]
\centering
\begin{tabular}{|l|c|}
 \hline
  &  \\[-2.4mm]
 Systematic term & $\Delta r_0$ \\[2mm]
 \hline
 \hline
  1. Jet finder dependence  & 0.043 \\ 
  2. Fit interval & 0.025 \\ 
  3. Parametrization of $\nee$  & 0.067 \\ %
  4. Value of $\Lambda$  & 0.024 \\ 
  5. Averaging procedure for $\qscale$ and $\gscale$ & 0.044 \\ 
  6. Number of active flavors & 0.002 \\ 
 \hline
 \hline
  Total  &  0.097 \\
 \hline
\end{tabular}
\caption{Summary of systematic uncertainties
for the effective value of $r_0$ at the scale of the~Z$^0$
as determined using data~\cite{bib-delphi99}.
}
\label{tab-systerr}
\end{table}


\newpage

\newpage

\newpage

\newpage

\newpage

\newpage


\newpage
\begin{thebibliography}{99}
\bibitem{bib-qcd}
  See, for example,
  R.K. Ellis, W.J. Stirling, B.R. Webber,
  ``QCD and Collider Physics,'' 
  Cambridge University Press (1996).
\bibitem{lep-4jet}
  OPAL Collaboration, R. Akers et al., Z. Phys. C {\bf 65}, 367 (1995); \\
  DELPHI Collaboration, P. Abreu et al., Phys. Lett. B {\bf 414}, 401 (1997); \\
  ALEPH Collaboration, R. Barate et al., Z. Phys. C {\bf 76}, 1 (1997).
\bibitem{bib-opal99}
  OPAL Collaboration, G. Abbiendi et al., CERN-EP/99-028,
  in press in Eur. Phys. J.~C.
\bibitem{bib-delphi99}
  DELPHI Collaboration, P. Abreu et al.,
  Phys. Lett. B {\bf 449}, 383 (1999).
\bibitem{bib-khoze}
  V.A. Khoze, W. Ochs, Int. J. Mod. Phys. A {\bf 12}, 2949 (1997).
\bibitem{bib-mueller}
  J.B. Gaffney and A.H. Mueller, Nucl. Phys. B {\bf 250}, 109 (1985).
\bibitem{bib-dokshitzer}
  Yu.L. Dokshitzer, V.A. Khoze, S.I. Troyan,
   Sov. J. Nucl. Phys. {\bf 47}, 881 (1988).
\bibitem{bib-capella}
  A. Capella, I.M. Dremin, J.W. Gary, V.A. Nechitailo, 
  J. Tran Thanh Van, hep-ph/9910226, 
  in press in Phys. Rev.~{\bf D}.
\bibitem{bib-dgary99}
   I.M. Dremin, J.W. Gary, hep-ph/9905477 v2,
   Phys. Lett. B {\bf 459}, 341 (1999).
\bibitem{bib-pdg}
   Review of Particle Physics,
   C. Caso et al., Eur. Phys. J. C {\bf 3}, 1 (1998).
\bibitem{bib-opal91}
  OPAL Collaboration, M.Z. Akrawy et al. Phys. Lett. B {\bf 263}, 311 (1991); \\
  OPAL Collaboration., G. Alexander et al., Phys. Lett. B {\bf 265}, 462 (1991).
\bibitem{bib-ezero}
  See, for example,
  OPAL Collaboration, M.Z. Akrawy et al., Z. Phys. C {\bf 49}, 375 (1991).
\bibitem{bib-herwig}
  G. Marchesini, B.R. Webber et al.,
  Comp. Phys. Comm. {\bf 67}, 465 (1992).
\bibitem{bib-opal96}
  OPAL Collaboration, G. Alexander et al.,
  Phys. Lett. B {\bf 388}, 659 (1996).
\bibitem{bib-jetset}
  T. Sj\"{o}strand, Comp. Phys. Comm. {\bf 82}, 74 (1994); \newline
  T. Sj\"{o}strand, CERN-TH.7112/93 (revised August 1995).
\bibitem{bib-opal96b}
   OPAL Collaboration, G. Alexander et al., Z. Phys. C {\bf 69}, 543 (1996).
\bibitem{bib-durham}
   S. Catani et al., Phys. Lett. B {\bf 269}, 432 (1991).
\bibitem{bib-jade}
   JADE Collaboration, W. Bartel et al., Z. Phys. C {\bf 33}, 23 (1986).
\bibitem{bib-cone}
   OPAL Collaboration, R. Akers et al., Z. Phys. C {\bf 63}, 197 (1994).
\bibitem{bib-dn}
   I.M. Dremin, V.A. Nechitailo, Mod. Phys. Lett. A {\bf 9}, 1471 (1994).
\bibitem{bib-nlo}
   B.R. Webber, Phys. Lett. B {\bf 143}, 501 (1984).
\bibitem{bib-opal97}
  OPAL Collaboration, K. Ackerstaff et al.,
  Z. Phys. C {\bf 75}, 193 (1997).
\end{thebibliography}
\end{document}